\documentclass[a4paper,11pt]{article}

\usepackage{jheppub,bm,color}
\newcommand{\beq}{\begin{eqnarray}}
\newcommand{\eeq}{\end{eqnarray}}
\renewcommand\d{\partial}

\begin{document}

\title{Axion crystals}

\author{Sho Ozaki and Naoki Yamamoto}
\affiliation{Department of Physics, Keio University, Yokohama 223-8522, Japan}
\affiliation{Research and Education Center for Natural Sciences, Keio University, Yokohama 223-8521, Japan}
\emailAdd{sho.ozaki@keio.jp}
\emailAdd{nyama@rk.phys.keio.ac.jp}

\abstract
{The low-energy effective theories for gapped insulators are classified by three parameters: 
permittivity $\epsilon$, permeability $\mu$, and theta angle $\theta$. Crystals with periodic 
$\epsilon$ are known as photonic crystals. We here study the band structure of photons in a new 
type of crystals with periodic $\theta$ (modulo $2\pi$) in space, which we call the axion crystals. 
We find that the axion crystals have a number of new properties that the usual photonic crystals 
do not possess, such as the helicity-dependent mass gap and nonrelativistic gapless dispersion 
relation at small momentum. We briefly discuss possible realizations of axion crystals in 
condensed matter systems and high-energy physics.}
\maketitle

\section{Introduction}
The electronic band structure that originates from the periodicity of crystals is essential in 
determining the electronic properties of materials. Yet the notion of the band structure is not 
limited to electrons, but can also be extended to photons: a regular array of materials with different 
dielectric constants (or different refractive indices) leads to the band structure of photons. 
% Here the periodic dielectric function for photons plays the same rule, in structure, as the periodic atomic potential for electrons. 
This is the photonic crystal \cite{Joannopoulos}, which is expected to be applied to various 
optoelectronic devices. 

Meanwhile, recent development of topological insulators (see, e.g., refs.~\cite{Hasan:2010xy, Qi:2011zya} 
for reviews) reveals that, most generically, the low-energy effective theories for all the gapped 
insulators are classified by not only permittivity $\epsilon$ and permeability $\mu$, but also one 
more parameter---the theta angle $\theta$. Here $\theta \equiv 0$ and $\theta \equiv \pi$ (modulo $2\pi$),
corresponding to normal insulators and topological insulators, are only possible in time-reversal 
systems \cite{Qi:2008ew}.

In this paper, we consider a new type of crystals with periodic $\theta$ (modulo $2\pi$) in space, 
which we call the \emph{axion crystals}. As the simplest yet nontrivial demonstration that allows 
for analytic treatment, we study the band structure of photons in one-dimensional axion crystals
with the staircase profile of $\theta$ (see figure~\ref{fig:profile} below). As we will see, this 
configuration corresponds to the ``Kronig-Penney-type potential" for $\bm{\nabla}\theta$.
 
We find that axion crystals have a number of new features that the usual photonic crystals do not
possess. In particular, one of the helicity states acquires a mass gap, even without superconductivity, 
while the other has the nonrelativistic gapless dispersion relation at small momentum. The presence 
of such gapped photon and nonrelativistic gapless photon depending on the helicity states in the 
axion electrodynamics has been recently pointed out in ref.~\cite{Yamamoto:2015maz} for 
homogeneous $\bm{\nabla}\theta$ .

This paper is organized as follows. In section~\ref{sec:eft}, we review the generic low-energy effective
theory for gapped insulators. We also derive the master equation of the photon for a given 
configuration of $\theta$. In section~\ref{sec:band}, we study the properties of the band structure
of one-dimensional axion crystals. Section \ref{sec:discussion} is devoted to discussions.
Throughout the paper, we use the units $\hbar = c = e =1$.

\section{Low-energy effective theory of insulators}
\label{sec:eft}
We first briefly review the generic low-energy effective theory of insulators. Because the electron 
has a mass gap in insulators by definition, the only relevant low-energy degrees of freedom are 
photons described by the gauge field $A^{\mu}$. The effective theory for a given insulator that 
respects the gauge symmetry and the rotational symmetry in space can be written down up to 
the second order in derivatives as \cite{Wilczek:1987mv}
\beq
\label{L_axion}
\mathcal{L}
= \frac{\epsilon}{2} \bm{E}^{2} - \frac{1}{2 \mu} \bm{B}^{2} + \frac{\theta}{4\pi^2} \bm{E} \cdot \bm{B} + A_{\mu} j^{\mu}\,.
\eeq
Here $\epsilon$, $\mu$, and $\theta$ are some constants that depend on the microscopic details 
of the system, and $j^{\mu} = (\rho, \bm{j})$ is the electric current density. The quantities $\epsilon$ 
and $\mu$ are called the permittivity and permeability, respectively, which characterize the violation 
of the Lorentz invariance in medium, and $\theta$ is called the axion term.
 
Under the time reversal symmetry (which will be abbreviated as $\cal{T}$ symmetry), the 
electromagnetic fields are transformed as $\bm{E} \rightarrow \bm{E}$ and $\bm{B} \rightarrow -\bm{B}$. 
Hence, $\cal{T}$ symmetry is broken for generic $\theta$. Meanwhile, the path integral quantization 
of the theory (\ref{L_axion}) shows that all the physical quantities are invariant under the transformation 
$\theta \rightarrow \theta + 2\pi$ (see ref.~\cite{Vazifeh:2010pq} and references therein). Thanks to this 
periodicity, $\theta = -\pi$ and $\theta = \pi$ are equivalent; the system respects $\cal{T}$ symmetry when 
\beq
\theta \equiv 0 \ \ {\rm or } \ \ \pi \ \ ({\rm mod} \ 2\pi).
\eeq
The insulators with $\theta \equiv 0$ (mod $2\pi$) are normal (or topologically trivial) insulators and 
those with $\theta \equiv \pi$ (mod $2\pi$) are topological insulators \cite{Qi:2008ew}. The modified 
electrodynamics with the spacetime dependent $\theta = \theta(t, \bm{x})$ is known as the 
axion electrodynamics \cite{Wilczek:1987mv}.

In the following, we shall consider nondispersive media described by the axion electrodynamics~(\ref{L_axion}) 
with the space-dependent $\theta$ term, $\theta = \theta(\bm{x})$, and constant (space-independent) 
$\epsilon$ and $\mu$ to demonstrate new features for photons induced by the $\theta$ term alone.
We will later consider some specific form of $\theta(\bm{x})$ below (see figure~\ref{fig:profile}), but for a 
moment we will keep our argument general without restricting the particular form of $\theta$.
More generally, one can consider the space-dependent $\epsilon$ and/or $\mu$, as usual photonic 
crystals, at the same time. One can also consider dispersive media where the electric polarization and 
magnetization are given by the linear convolutions of the response functions multiplied by the 
electric and magnetic fields. Such extensions should be straightforward and are left for future work. 

The modified Maxwell's equations in the presence of $\theta(\bm{x})$ that follow from the 
Lagrangian (\ref{L_axion}) are given by \cite{Wilczek:1987mv}
\begin{gather}
\label{Gauss}
\epsilon \bm{\nabla} \cdot \bm{E} = \rho - \frac{1}{4\pi^2} \bm{\nabla} \theta \cdot \bm{B} \,, \\
\label{Ampere}
\frac{1}{\mu} \bm{\nabla} \times \bm{B} 
=  \epsilon \d_t {\bm E} + \bm{j} + \frac{1}{4\pi^2} \bm {\nabla} \theta \times \bm{E} \,, \\
\label{no_monopole}
\bm{\nabla} \cdot \bm{B} = 0\,, \\
\label{Faraday}
\d_t \bm{B} = - \bm{\nabla} \times \bm{E}\,.
\end{gather}
The additional contributions to the usual Maxwell's equations are the anomalous charge in the 
second term on the right-hand side of eq.~(\ref{Gauss}) and the anomalous Hall effect in the 
third term on the right-hand side of eq.~(\ref{Ampere}).

\section*{Master equation in the axion electrodynamics}
Let us consider the propagation of an electromagnetic wave (or a photon) in this type of medium 
without the source of photons, i.e., $\rho = 0$ and $\bm{j} = {\bm 0}$.
From eqs.~(\ref{Ampere}) and (\ref{Faraday}), we get the closed equation for ${\bm E}$,
\beq
\partial_{t}^{2} \bm{E}
&=& v^{2} \bm{\nabla}^{2} \bm{E} - \frac{1}{4\pi^2 \epsilon} \bm{\nabla} \theta \times \d_t \bm{E}\,,
\eeq
where $v = 1/\sqrt{ \epsilon \mu}$ is the velocity of the photon in medium.

We assume that $\theta({\bm x})$ has the $z$ dependence and consider the photon propagating in the 
$z$ direction. Then we find the following coupled equations for $E_x$ and $E_y$:
\begin{align}
\d_t^2 E_{x} (t, z) - v^{2} \d_z^2 E_{x} (t, z) - \frac{\d_z \theta(z)}{4\pi^2 \epsilon} \d_t E_{y} (t, z) &= 0 \,, \\
\d_t^2 E_{y} (t, z) - v^{2} \d_z^2 E_{y} (t, z) + \frac{\d_z \theta(z)}{4\pi^2 \epsilon} \d_t E_{x} (t, z) &= 0 \,.
\end{align}
Since $E_{y} = \pm i E_{x}$ for the positive and negative helicity states of the photon ($h = \pm1$), these 
equations become \cite{Yamamoto:2015maz}
\beq
\d_t^2 E_{x} (t, z) - v^{2} \d_z^2 E_{x} (t, z) 
\mp i \frac{\d_z \theta(z)}{4\pi^2 \epsilon} \d_t E_{x} (t, z) = 0 \qquad (h = \pm1)\,.
\eeq
The temporal translational symmetry allows us to take the Fourier transform with respect to time as
\beq
\label{E}
\omega^{2} E_{x}(z) + v^{2} \d_z^2 E_x(z) \pm \frac{\d_z \theta(z) }{4\pi^2 \epsilon} \omega E_{x} (z) = 0 \qquad (h = \pm1)\,,
\eeq
where $\omega$ is the frequency. This is the master equation that determines the behavior of $E_x(z)$
(and similarly that of $E_y(z)$) for a given configuration of $\d_z \theta(z)$.

\section{Band structure in axion crystals}
\label{sec:band}
\begin{figure}
\begin{minipage}{1.0\hsize}
\begin{center}
\includegraphics[width=0.9 \textwidth]{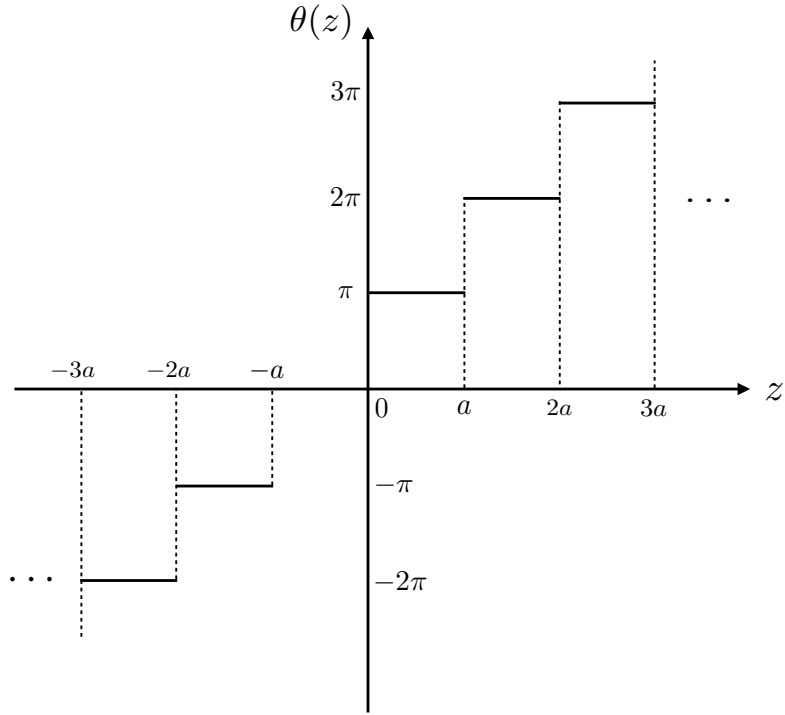}
\end{center}
\end{minipage}
\caption{Profile of $\theta(z)$.
}\label{fig:profile}
\end{figure}

\begin{figure}
\begin{minipage}{1.0\hsize}
\begin{center}
\includegraphics[width=1.0 \textwidth]{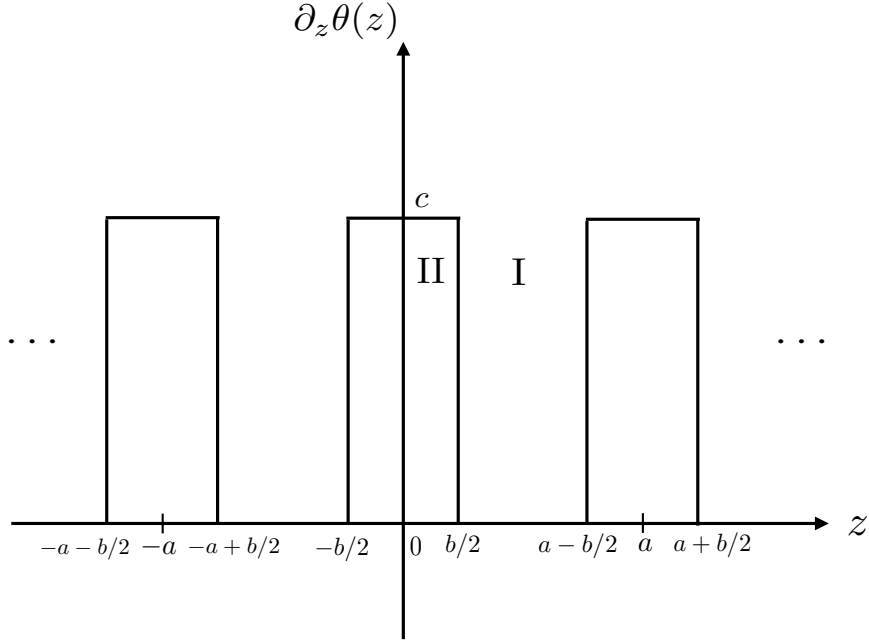}
\end{center}
\vspace{-2cm}
\end{minipage}
\caption{``Kronig-Penney-type potential" for $\d_z \theta(z)$.}
\label{fig:Kronig-Penney}
\end{figure}

We now study the band structure (or dispersion relation) of photons in the axion electrodynamics
with periodic $\d_z \theta(z)$ in space. Formally, this is analogous to the calculation of the band 
structure of electrons by solving the Schr\"odinger equation in a periodic potential.
As the simplest yet nontrivial example which allows for analytical treatment, we here consider the 
one-dimensional ``Kronig-Penney-type potential" for $\d_z \theta(z)$ as depicted in 
figure~\ref{fig:Kronig-Penney}. Later we will take the limit $b \to 0$ and $c \to \infty$ such that 
$b  c = \pi$. In this case, the ``potential" $\d_z \theta(z)$ is expressed as
\beq
\d_z \theta(z) = \pi \sum_{n=-\infty}^{\infty} \delta(z - na)\,.
\eeq
This corresponds to the staircase configuration of $\theta(z)$ shown in figure~\ref{fig:profile}, which
can be regarded as a regular array of trivial insulators with $\theta = 2n \pi$ ($n \in \mathbb{Z}$) 
and topological insulators with $\theta = (2n+1) \pi$ (with the same $\epsilon$ and $\mu$) with lattice 
constant $a$ in one direction.%
\footnote{Strictly speaking, without any ${\cal T}$ symmetry breaking, there would be gapless 
Dirac fermions at the interfaces between trivial and topological insulators, which would invalidate 
the description of the pure axion electrodynamics above. Here our description of the pure axion 
electrodynamics should be understood as being valid below the gap of Dirac fermions in the 
presence of a ${\cal T}$ symmetry breaking perturbation at the interfaces \cite{Qi:2008ew}. See 
also ref.~\cite{Casimir} for a similar setup in the context different from the band structure.} 
Note that this configuration of $\theta$ is periodic modulo $2\pi$. 
We call this type of systems the \emph{axion crystals}.

%%%%%%%%%%%%%%%%%%%%%%%%%
%%%%%%%%%%%%%%%%%%%%%%%%%

%%%%%%%%%%%%%%%%%%%%%%%%%
%%%%%%%%%%%%%%%%%%%%%%%%%

Let us first consider the photon with the helicity state $h=+1$.
In the region I ($b/2 < z < a-b/2$) where $\partial_{z} \theta(z)=0$, eq.~(\ref{E}) becomes
\beq
\omega^{2} E_x (z) + v^{2} \d_z^2 E_{x}(z) = 0.
\eeq
The generic solution is given by
\beq
E_{x}(z) = A {\rm{e}}^{-iKz} + B {\rm{e}}^{iKz}, \qquad K \equiv \frac{\omega}{v} \,,
\eeq
with some constants $A$ and $B$.

In the region II ($-b/2 < z < b/2$) where $\partial_{z} \theta(z)=c$, eq.~(\ref{E}) is
\beq
\omega^{2} E_{x}(z) + v^{2} \d_z^2 E_{x}(z) + \frac{c}{4\pi^2 \epsilon} \omega E_{x}(z) = 0\,.
\eeq
The solution is given by
\beq
E_{x}(z) = C {\rm{e}}^{-iQz} + D {\rm{e}}^{iQz}, \qquad
Q \equiv \frac{\omega}{v} \left(1 + \frac{c}{4\pi^2 \epsilon \omega } \right)^{\! 1/2} \,,
\eeq
with some constants $C$ and $D$.

We now impose the boundary conditions at $z=b/2$. Integrating eq.~(\ref{E}) over the interval 
$[b/2-\delta, b/2+\delta]$ and taking the limit $\delta \rightarrow 0$, it follows that $\d_z E_x(z)$ 
together with $E_{x}(z)$ must be continuous at $z=b/2$. We thus obtain
\beq
\label{condition1}
A {\rm{e}}^{-iKb/2} + B {\rm{e}}^{iKb/2} &=&  C {\rm{e}}^{- iQb/2} + D {\rm{e}}^{ iQb/2}, \\
\label{condition2}
-iKA {\rm{e}}^{-iKb/2} + iKB {\rm{e}}^{iKb/2} &=& -iQ C {\rm{e}}^{-iQb/2 } + iQ D {\rm{e}}^{i Qb/2 }.
\eeq
In addition, as $\d_z \theta(z)$ in eq.~(\ref{E}) is periodic with the period $a$, Bloch's theorem states that 
$E_x(z)$ satisfies $E_x (z+a) = e^{ika} E_x (z)$, and consequently, $\d_z E_x (z+a) = e^{ika} \d_z E_x (z)$. 
These conditions at $z=-b/2$ lead to
\beq
\label{condition3}
A {\rm{e}}^{-iK(a-b/2) } + B {\rm{e}}^{iK (a-b/2) } &=& \left(C {\rm{e}}^{i Qb/2 } + D {\rm{e}}^{-i Qb/2 } \right) {\rm{e}}^{ika}, \\
\label{condition4}
-iK A {\rm{e}}^{-iK(a-b/2) } + iKB {\rm{e}}^{iK (a-b/2) } &=& \left( -iQC {\rm{e}}^{i Qb/2 } + iQD {\rm{e}}^{-i Qb/2 } \right) {\rm{e}}^{ika}.
\eeq
Equations (\ref{condition1})--(\ref{condition4}) can be summarized in the matrix equation, 
${\cal M}_{i j} x^j =0$, where
\beq
{\cal M} \equiv \!
\left(
\begin{tabular}{cccc}
${\rm{e}}^{-iKb/2}$ & ${\rm{e}}^{iKb/2}$ & $-{\rm{e}}^{-iQb/2}$ & $-{\rm{e}}^{iQb/2}$ \\
$-iK{\rm{e}}^{-iKb/2}$ & $iK{\rm{e}}^{iKb/2}$ & $iQ{\rm{e}}^{-iQb/2}$ & $-iQ{\rm{e}}^{iQb/2}$ \\
${\rm{e}}^{-iK(a-b/2)}$ & ${\rm{e}}^{iK(a-b/2)}$ & $-{\rm{e}}^{iQb/2 + ika}$ & $-{\rm{e}}^{-iQb/2+ ika}$ \\
$-iK{\rm{e}}^{-iK(a-b/2)}$ & $iK{\rm{e}}^{iK(a-b/2)}$ & $iQ{\rm{e}}^{iQb/2 + ika}$ & $-iQ{\rm{e}}^{-iQb/2+ ika}$ 
\end{tabular}
\right)\,\!,
\quad \! \!
{\bm x} \equiv \! \left(
\begin{tabular}{c}
$A$ \\
$B$ \\
$C$ \\
$D$
\end{tabular}
\right)\,\!.
\nonumber \\
\eeq
It has a nontrivial solution as long as the condition, $\det {\cal M} = 0$, is satisfied.

This condition is simplified in the limit $b \to 0$ and $c \to \infty$ with keeping $bc = \pi$, 
corresponding to the profile of $\theta(z)$ in figure~\ref{fig:profile}.
After a straightforward calculation, we get
\beq
\label{condition+}
\cos(ka) = \cos(Ka) - \frac{1}{8\pi} \sqrt{\frac{\mu}{ \epsilon}} \sin(Ka)\,,
\eeq
for $h=+1$. Similarly, we have
\beq
\label{condition-}
\cos(ka) = \cos(Ka) + \frac{1}{8\pi} \sqrt{\frac{\mu}{ \epsilon}} \sin(Ka)\,,
\eeq
for $h=-1$. The second terms on the right-hand sides of eqs.~(\ref{condition+}) and (\ref{condition-})
originate from the $\theta$ term, which distinguishes the two helicity states of photons.

Without the $\theta$ term, two equations above become degenerate, $\cos(ka) = \cos(Ka)$, 
reducing to the usual dispersion relation of the photon in medium,
\beq
\omega = v k.
\eeq
However, the presence of the periodic $\theta$ term gives rise to a nontrivial band structure 
of photons, analogously to that of electrons in a periodic potential, and further to a band structure
splitting between the two helicity states of photons as a consequence of the parity-violating nature 
of the $\theta$ term, unlike usual photonic crystals.

\begin{figure}[h]
\begin{minipage}{1.0\hsize}
\begin{center}
\includegraphics[width=1.0 \textwidth]{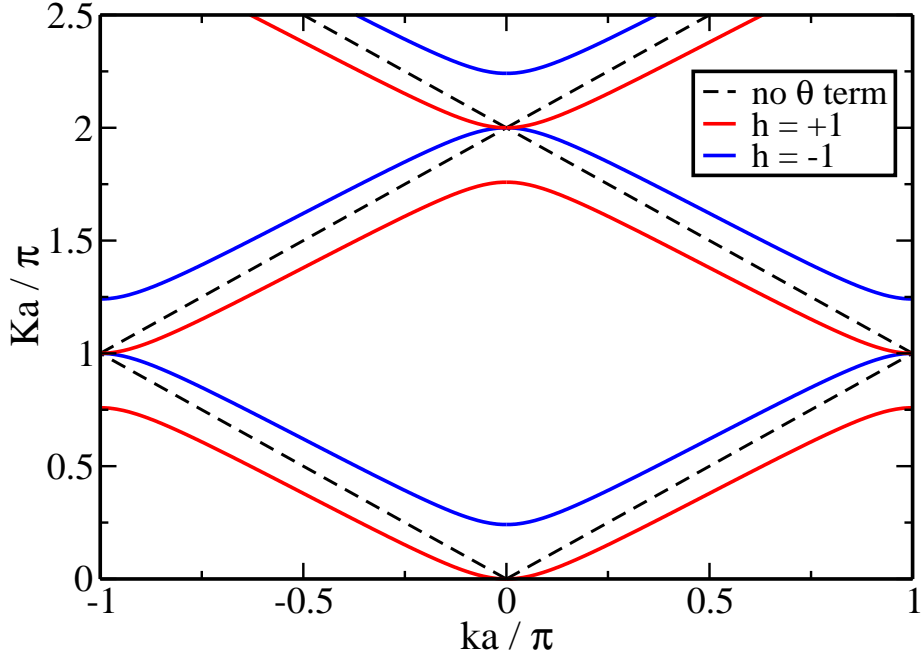}
\vskip -0.1in
\end{center}
\end{minipage}
\caption{Band structures of photons for $h = \pm 1$ without (black dashed line) 
and with (red and blue lines) the $\theta$ term. The figure here is shown for 
$\sqrt{\mu/\epsilon}=10$ as an example.}
\label{fig:band}
\end{figure}

\begin{figure}[h]
\begin{minipage}{1.0\hsize}
\begin{center}
\includegraphics[width=0.9 \textwidth]{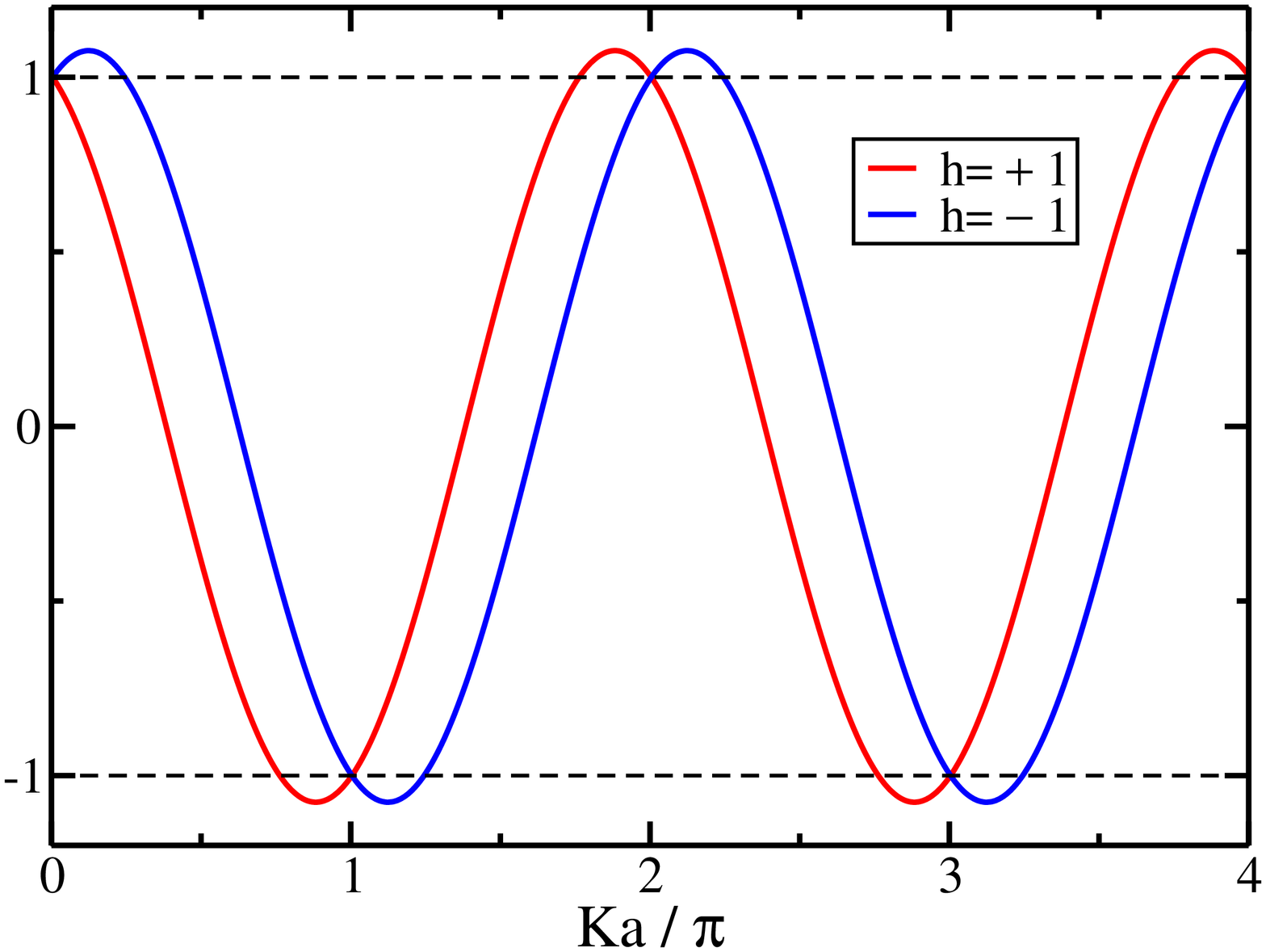}
\vskip -0.1in
\end{center}
\end{minipage}
\caption{
The right-hand sides of eqs.~(\ref{condition+}) and (\ref{condition-}) as a function of $Ka$. 
}\label{fig:rhs}
\end{figure}

From eqs.~(\ref{condition+}) and (\ref{condition-}), we plot the band structure of photons with 
each helicity $h= \pm 1$ in figure~\ref{fig:band}. One observes that, for each helicity, there is a 
range of frequency---the \emph{photonic band gaps} at $ka/\pi = n$ ($n=0,1,2,\cdots$)---where 
no photon eigenmode exists. This feature is similar to that in usual photonic crystals. To understand 
the origin of these band gaps, we plot the right-hand sides of the eqs.~(\ref{condition+}) and 
(\ref{condition-}) as a function of $Ka$ in figure~\ref{fig:rhs}. Since the left-hand sides of both 
equations is bounded as $-1 \leq \cos(ka) \leq +1$, the solutions do not exist when the right-hand 
sides are outside of this region for some $Ka$. The absence of the solutions suggests the 
presence of the photonic band gaps for such $Ka$. 

%%%%%%%%%%%%%%%%%%%%%%%%%
%%%%%%%%%%%%%%%%%%%%%%%%%

%%%%%%%%%%%%%%%%%%%%%%%%%
%%%%%%%%%%%%%%%%%%%%%%%%%

Using eqs.~(\ref{condition+}) and (\ref{condition-}), one can show that the magnitudes of the 
photonic band gaps for the helicity $h=\pm1$ at $ka/\pi = n$, which we define as $\Delta_n^{\pm}>0$, 
are all equivalent, except for $h=+1$ and $n=0$ (for which the photonic band gap is absent). 
Indeed, such $\Delta_n^{\pm}$ satisfies the same equation,
\beq
\label{eq:Delta}
\cos \left(\frac{\Delta_n^{\pm} a}{v} \right) + \frac{1}{8\pi} \sqrt{\frac{\mu}{\epsilon}} \sin \left(\frac{\Delta_n^{\pm} a}{v} \right) = 1\,,
\eeq
independently of $n$ and the helicity $h=\pm1$ (except for $h=+1$ and $n=0$), 
and one can set $\Delta_n^{\pm} = \Delta$.
Equation (\ref{eq:Delta}) can be easily solved in terms of $\Delta$ as 
\beq
\label{Delta}
\Delta = \frac{v}{a} \arcsin \left(\frac{2p}{p^2 + 1} \right)\,, \qquad p \equiv \frac{1}{8\pi} \sqrt{\frac{\mu}{\epsilon}}\,.
\eeq
Note that $0 \leq 2p/(p^2+1) \leq 1$ for any $p>0$.

In summary, we obtain the photonic band gaps for each helicity $h=\pm1$ as follows:
\begin{align}
\frac{n \pi v}{a} \leq \omega \leq \frac{n \pi v}{a} + \Delta \quad {\rm for} \ \ h=-1 \ \ (n=0,1,\cdots)\,, 
\nonumber \\
\frac{n \pi v}{a} - \Delta \leq \omega \leq \frac{n \pi v}{a} \quad {\rm for} \ \ h=+1 \ \ (n=1,2,\cdots)\,, 
\end{align}
where $\Delta$ is given by eq.~(\ref{Delta}).

It is interesting to note that there exists a special value $p=1$, or  
\beq
\frac{\mu}{\epsilon} = (8\pi)^2\,,
\eeq
where the photonic band gaps are always open for either only one of the two helicity states
of photons. In this case, the photonic band gaps are
\begin{align}
\frac{n \pi v}{a} \leq \omega \leq (2n+1)\frac{\pi v}{2a} \quad {\rm for} \ \ h=-1 \ \ (n=0,1,\cdots)\,,
\nonumber \\
(2n-1)\frac{\pi v}{2a} \leq \omega \leq \frac{n \pi v}{a} \quad {\rm for} \ \ h=+1 \  \ (n=1,2,\cdots)\,. 
\end{align}

\section*{Helical gapped photons and nonrelativistic photons}
Let us closely look at the properties of the band structure of photons at small $k$ in axion crystals.

First, as we have seen above, one of the helicity states ($h=-1$) is gapped at $k=0$. Remarkably,
the photon here acquires a mass gap even in the absence of superconductivity or the so-called 
Englert-Brout-Higgs mechanism. While mass gap generation for gauge fields  due to the 
Chern-Simons term in $2+1$ spacetime dimensions has been known \cite{Deser:1982vy}, our result
provides a new mechanism of {\it helicity-dependent} mass gap generation due to the axion term 
in $3+1$ dimensions.

Second, the other gapless helicity state ($h=+1$) does \emph{not} have a relativistic dispersion relation.
By performing the expansion in small $k$ and $\omega$ in eq.~(\ref{condition+}), we obtain
\beq
\omega = \frac{4\pi a}{\mu} k^2 + O(k^3)\,.
\eeq
Just like the usual relativistic photon with the dispersion relation $\omega = k$ can be regarded as 
the (type-I) Nambu-Goldstone mode \cite{Ferrari:1971at, Brandt:1974jw}, this ``nonrelativistic photon" 
may be understood as the so-called type-II Nambu-Goldstone mode \cite{Yamamoto:2015maz, Hidaka}.

The appearances of the gapped photon and nonrelativistic gapless photon depending on the 
helicity states are the novel features that are absent in usual photonic crystals.

\section{Discussions}
\label{sec:discussion}
In this paper, we have studied the band structure of photons in axion crystals. Although we have 
focused here on a one-dimensional axion crystal as a demonstration, one can consider axion 
crystals in two and three dimensions. Our simple model could also be made more realistic by 
relaxing the condition of the homogeneous permittivity and/or permeability. 

The axion crystal should be realized by stacking trivial insulators with $\theta = 2n \pi$ 
($n \in \mathbb{Z}$) and topological insulators with $\theta = (2n+1) \pi$ alternately.
The helicity-dependent band structure of the axion crystal suggests that electromagnetic waves 
with one of the circularly polarizations and with frequencies within the photonic band gaps 
incident in the direction of the modulation of $\theta$ are perfectly reflected. Hence, the axion 
crystal behaves as a polarizer that transmits electromagnetic waves with only one polarization 
within these frequency regions. One could, in principle, tune the region and the magnitude of 
the photonic band gaps by changing the lattice constant $a$ [see eq.~(\ref{Delta})]. 
This property may be useful for applications in possible new optoelectronic devices.

The axion crystals may be relevant in high-energy physics as well. It has been recently found that 
the ground state of Quantum Chromo Dynamics (QCD) at finite baryon chemical potential in a 
strong magnetic field becomes the \emph{chiral soliton lattice} of neutral pions, characterized by 
the periodic $\langle \bm{\nabla} \pi^0 \rangle$ in space \cite{Brauner:2016pko}. Then the 
Wess-Zumino-Witten term $\pi^0 \bm{E} \cdot \bm{B}$ leads to the axion electrodynamics in 
QCD matter \cite{Yamamoto:2015maz} (see also ref.~\cite{Ferrer:2015iop}), where the periodic 
$\langle \bm{\nabla} \pi^0 \rangle$ exactly plays the role of the periodic $\bm{\nabla} \theta$ 
considered in this paper. In this way, the chiral soliton lattice in QCD provides an explicit 
realization of the axion crystal. Such a state is potentially realized in dense matter inside neutron 
stars with a strong magnetic field.

\section*{Acknowledgement}
The author N.~Y. thanks T.~Brauner and Y.~Hidaka for useful conversations. 
This work was supported by JSPS KAKENHI Grant No.~16K17703 and MEXT-Supported Program 
for the Strategic Research Foundation at Private Universities, ``Topological Science" (Grant No.~S1511006).

\end{document}